# Amplified linear and nonlinear chiral sensing assisted by anapole modes in hybrid metasurfaces

Guillermo Serrera, Javier González-Colsa and Pablo Albella[a]

**AFFILIATION**

*Group of Optics, Department of Applied Physics, University of Cantabria, 39005, Spain.*
a) Author to whom correspondence should be addressed: *pablo.albella@unican.es*

**ABSTRACT**

The interaction between chiral molecules and circularly polarized light is largely influenced by the local optical chirality density. This interaction prompts substantial demand of the design of nanophotonic platforms capable of enhancing such effects across large and accessible volumes. Such a magnification requires nanostructures that provide strong electric and magnetic field enhancements while preserving the phase relation of circular light. Dielectric nanostructures, particularly those able to support resonances, are ideal candidates for this task due to their capacity for high electric and magnetic field enhancements. On the other hand, efficient third harmonic generation calls for strong electric field resonances within dielectric materials, a feature often boosted by incorporating plasmonic materials into hybrid systems. In this work, we numerically propose a coupled silicon disk-gold ring system that can exploit the anapole-induced field confinement to provide a broadband magnified circular dichroism under realistic conditions, reaching values up to a 230-fold enhancement. We also demonstrate that this structure can be employed as an efficient third harmonic generator which, when integrated with chiral media, enables an 800-fold enhancement in circular dichroism. Furthermore, we show that pulsed illumination at intensities up to 10 GW/cm$^2$ does not induce temperature increments that could potentially damage the samples. These findings suggest that this system can be a promising and versatile approach towards ultrasensitive chiral sensing.

Efficient discrimination of molecular chirality has lately emerged as a critical challenge for the nanophotonics community[1–5]. Chirality refers to the property whereby an object cannot be superimposed onto its mirror image[6–8], meaning that chiral molecules have mirror-image counterparts, known as enantiomers[9].

Although these enantiomers often share similar properties, their chirality plays a vital role in biological interactions[10,11]. Many bioreceptors are chiral, so the coupling between chiral molecules and bioreceptors is dependent on this handedness, which can lead to different outcomes. For instance, changes in protein chirality are likely responsible for serious health issues such as Parkinson and Alzheimer's diseases[12]. Thus, there is a growing interest in the precise enantiomer identification, especially in the biosciences and the pharmaceutical industry[13], where such distinction becomes critical.

Traditionally, enantiomer discrimination relied on Circular Dichroism (CD) spectroscopies. These techniques measure the differences in enantiomer absorption under right- and left-handed Circularly Polarized Light (CPL) illumination[9]. However, since these differences are intrinsically minuscule ($10^{-2} - 10^{-6}$ times the strength of absorption), this approach requires high sample concentrations and long measure periods to be effective[14]. This challenge has prompted many nanophotonic platforms to overcome these limitations.

As the authors Tang and Cohen demonstrated in their seminal work[15,16], the CD effect, expressed as the difference in absorption rates $A$ between the two circular polarizations can be calculated as

$$CD = A^R - A^L \propto \mathfrak{Im}(\kappa)C \qquad (1)$$

where $\kappa$ is the Pasteur parameter of the chiral medium, accounting for the coupling between electric and magnetic dipoles through the constitutive relations $\boldsymbol{D} = \varepsilon\boldsymbol{E} - i\kappa\sqrt{\varepsilon_0\mu_0}\boldsymbol{H}$ and $\boldsymbol{B} = \mu\boldsymbol{H} + i\kappa\sqrt{\varepsilon_0\mu_0}\boldsymbol{E}$; and $C$ is the Optical Chirality Density (OCD), defined as

$$C = -\frac{\omega\varepsilon_0}{2}\mathfrak{Im}(\boldsymbol{E}^* \cdot \boldsymbol{B}) = -\frac{\omega}{2c^2}|\boldsymbol{E}||\boldsymbol{H}|\cos(\Phi) \qquad (2)$$

where $\Phi$ is the phase angle between $i\boldsymbol{E}$ and $\boldsymbol{H}$. As can be seen from the right-hand side in equation 2, for fixed amplitude values of the electric and magnetic fields, OCD is maximized when the fields are collinear and the phase angle $\Phi$ is zero, equivalent to a $\pi/2$ phase difference[17]. For plane waves in vacuum, these conditions are both met in circular polarizations, maximizing the OCD to $C_{CPL} = \pm\frac{\omega\varepsilon_0}{2c}|\boldsymbol{E}|^2$. Furthermore, with the use of nanophotonic structures, enhanced optical fields can be engineered to obtain higher values of the OCD if the aforementioned conditions are satisfied[17].

Several approaches to improve chirality detection have come from plasmonics, exploiting the intense local fields created by Localized Surface Plasmon Resonances (LSPRs) nearby structures[18]. These resonances induce strong local OCD enhancements, but spatial variations of its sign are often found, leading to negligible spatial averages[19]. Moreover, LSPR excitation in metals, which have resistive losses, usually translates into thermal effects that might damage samples[20,21].

On the other hand, OCD enhancement with High Refractive Index Dielectric (HRID) structures is achieved based on both electric and magnetic field enhancements through Mie resonances[22–24]. This is particularly relevant when these spectrally overlap, as it occurs at the Kerker conditions[25,26] or in dielectric nanodisks, where resonances can be tuned by manipulating the geometrical aspect ratio[27,28]. Thus, dual structures with both metal and dielectric materials have led to optimal OCD enhancements[29].

Interestingly, HRID structures can also support anapole resonances, characterized by non-radiation and extreme field confinement[30]. Anapole modes have been relevant for several applications in nanophotonics[31], such as invisibility effects[32] or nanolasing[33]. In chiral sensing, anapolar field confinement has been speculated to provide background-free OCD enhancements[34]. This has also been exploited for boosting nonlinear optical effects, particularly in Third Harmonic Generation (THG)[35,36]. Specifically, the usage of HRID materials together with metallic materials has led to very efficient THG in hybrid metastructures[37,38].

In this work, we numerically show how a hybrid structure, composed of an HRID (amorphous silicon) holed nanodisk (consisting of a small hollow cylindrical gap located in the center of the silicon disk) coupled to a gold ring shows as a promising platform for chiral sensing, providing a broadband high average OCD value across a significant volume. This enhancement comes from the confinement of electric and magnetic energy near anapole resonances, which can also be exploited to provide an enhanced nonlinear Third Harmonic (TH) CD signal. Moreover, we demonstrate its high thermal tolerance to the illumination conditions needed to generate a significant TH signal without inflicting thermal damage to the sample.

Dielectric disks offer the possibility of tuning their resonances by changing the radius/height aspect ratio. To evaluate the potential of these resonances to enhance chirality, and following the strategy shown in[34], we perform 3D finite-difference time-domain (FDTD) numerical simulations on amorphous silicon holed disks (material data from Palik[39]), embedded in water ($n_{Wa} = 1.33$), closely resembling conditions in real experiments[40]. The choice of amorphous instead of crystalline silicon is based on its nonlinear properties, as amorphous silicon shows significantly lower two-photon absorption while keeping similar optical properties in the infrared range[41].

We set the disk and inner gap radii, $R_{Si} = 300$ nm and $r_{Si} = 25$ nm; and explore disk heights from $h = 150$ nm to $h = 510$ nm (corresponding to aspect ratios $h/R_{Si} = 0.5 - 1.7$ in steps of 0.1). In these simulations, the electric and magnetic fields, and OCD enhancement $\hat{C}$ within the inner gap are calculated and spatially averaged. Since the mesh size within the gap was uniform, the averaging procedure consisted in a simple arithmetic mean. A far-field multipole decomposition[42,43] was performed to obtain insight into the evolution of the different resonances with respect to the aspect ratio. These results are summarized in Figure 1. Scattering cross section results, together with their respective multipole reconstruction, can be found in the Supplementary Material (Figure S1).

Generally, there is a high OCD area (Figure 1a) following the Magnetic Dipole (MD) resonance, owing to high magnetic field enhancement at that resonance (Figure 1c), together with the Electric Dipole (ED), which offers a strengthening of the electric field (Figure 1b). This joint boost of both fields and the phase condition of circular light $\cos(\Phi)$ satisfying equation 2 lead to significant OCD enhancement. Additionally, the Electric Dipole Anapole (EDA) mode shows a notable trace of local electric field enhancement, as seen in Figure 1b. This finding, and the preservation of the phase condition (Figure 1d), set the basis for our proposal.

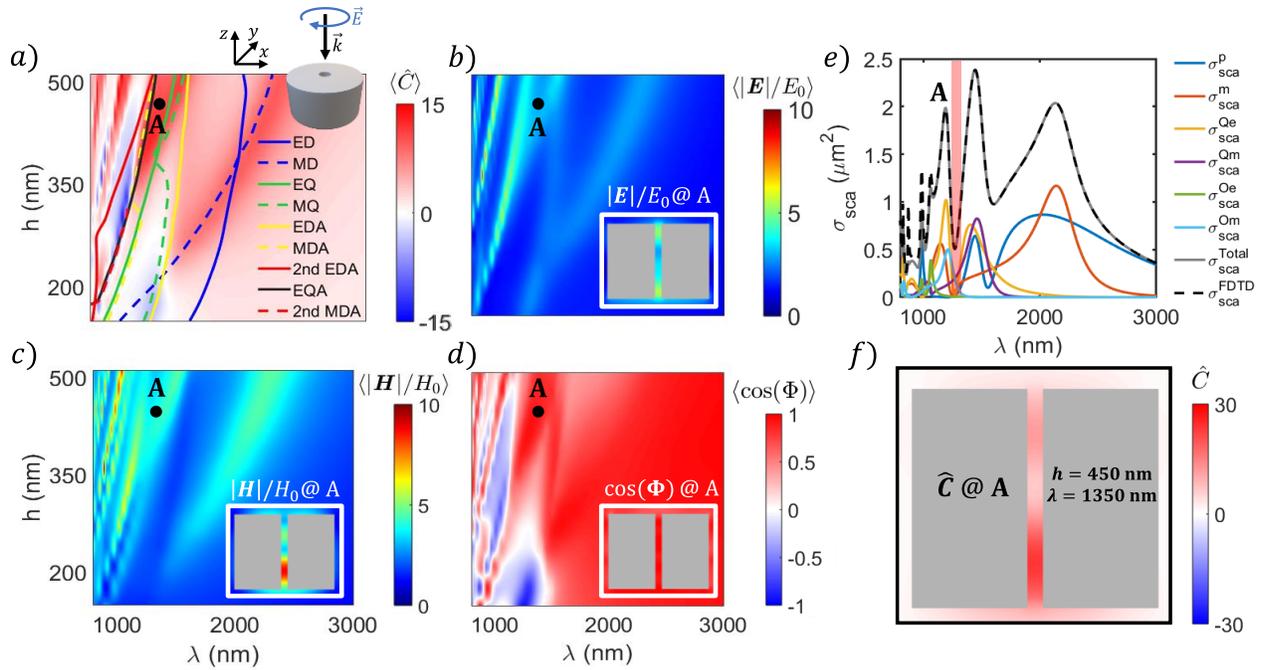

Figure 1. Evolution of the a) OCD, b) electric field, and c) magnetic field enhancements, and d) phase factor $\cos(\Phi)$ averaged within the inner gap with respect to the disk height and the incident wavelength $\lambda$. In a), marked with solid lines, the evolution of the different resonances can be seen. Insets with spatial distributions in the central $xz$ plane at point A ($h = 450$ nm, $\lambda = 1350$ nm) are shown in b)-d). e) Far field scattering cross-section for the $h = 450$ nm case. In red, the spectral position of point A, where anapole resonances overlap, is highlighted. f) Spatial distribution of the OCD (central $xz$ plane) enhancement at point A.

The EDA mode, although useful for several purposes, such as nonlinear effect boosting, has a limited utility by itself in terms of OCD. However, when mixed with other strong magnetic resonances, it can provide remarkable conditions for both purposes. This is what is found in the (1200-1400 nm, 350-500 nm) region in Figure 1a, with strong OCD averages (up to 15). In this region, there is a significant overlap of several modes, notably the Electric Quadrupole (EQ), its Anapole (EQA), the Magnetic Quadrupole (MQ) and a Magnetic Dipole Anapole (MDA) mode (see Figure 1e). This superposition creates an intense concentration of electric and magnetic fields, which together with a conservation of phase (as seen in Figure 1d), leads to a very strong average OCD. As it can be seen in Figure 1f, although this area shows an inhomogeneous OCD enhancement, owing to an uneven concentration of magnetic field in the gap, the OCD enhancement sign is unfluctuating.

Calculations for a more realistic system, i.e., considering a high thermal conductivity substrate (amorphous alumina, $n_{Al} = 1.75$) can be found in the Supplementary Material (Figure S2). There, the addition of a substrate causes a mild reduction in the overall enhancement, but the same trends can still be found. For the next step, which consists of adding a gold ring that enhances the nonlinear response while keeping a similar chiroptical response, we choose an optimal size of $h = 450$ nm (aspect ratio $h/R = 1.5$).

After finding an optimal silicon disk configuration, we add gold rings (material properties from Johnson & Christy[44]) to enhance the nonlinear response of the HRID disk while preserving the OCD enhancement. To find an optimal configuration, we explore rings with inner radii $r_{Au} \in [350,550]$ nm, where the minimum value corresponds to a separation of 50 nm from the silicon disk; and outer radii $R_{Au} \in [400,750]$ nm. The optimization process consisted on first fixing the inner radius to its minimum value ($r_{Au} = 350$ nm) and changing the outer radius until an optimum point was found. Then, the same optimization process was applied to the inner radius for the optimal outer radius. The ring height was fixed to $t_{Au} = 100$ nm.

The results for the optimization can be found in Figure 2a, where we can see that a high average OCD band (values close to 15) over a relatively broad wavelength range covering from around 1230 nm to 1500 nm appears. In particular, the outer radius does not affect significantly to the overall chirality within this band.

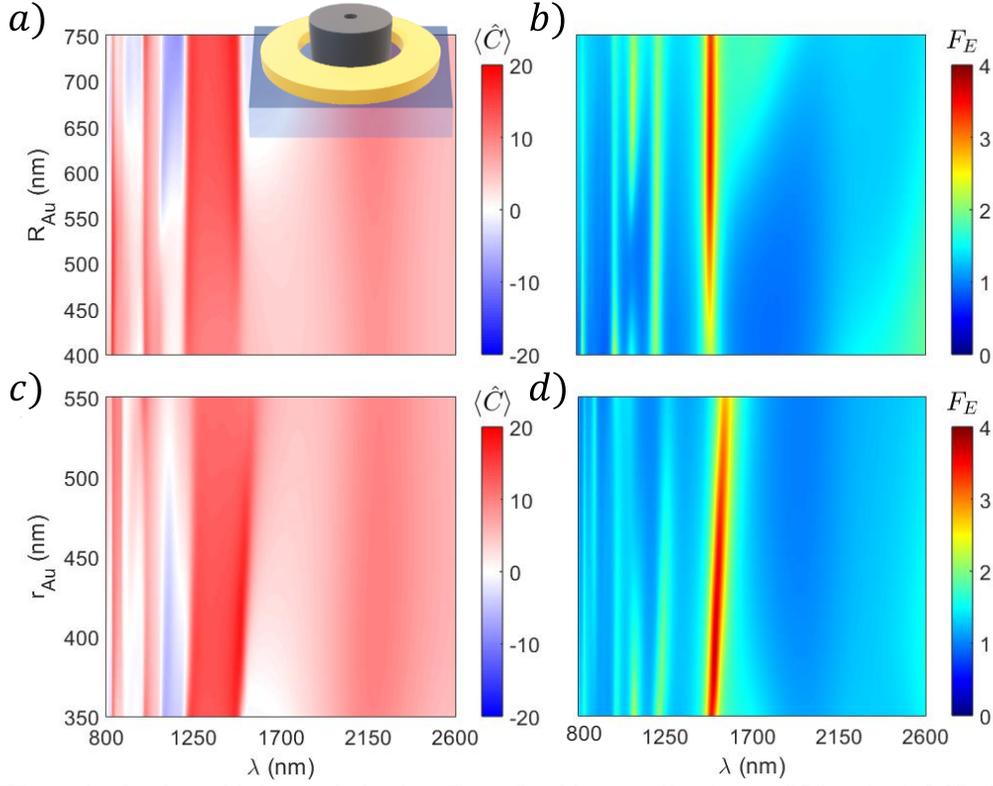

Figure 2. FDTD results for the gold ring optimization. For a fixed inner radius ($r_{Au} = 350$ nm): a) OCD, b) $F_E$ factor inside the silicon disk. For a fixed outer radius ($R_{Au} = 650$ nm): c) OCD, d) $F_E$ factor.

Since the geometry of the gold ring does not seem to significantly affect the chiral properties of the overall structure, we instead focus on optimizing its potential for THG. In order to predict its capabilities, the electric field intensity confined inside the silicon disk is calculated, giving rise to the $F_E$ factor[37]:

$$F_E = \frac{\iiint |\boldsymbol{E}|^2 \, dV}{|E_0|^2 V} \qquad (3)$$

where $V$ is the volume of the Si nanodisk. In nanophotonic structures, THG is optimized by maximizing the excitation density, i.e., the electric field intensity. This means that, we can predict efficient THG from high values in the $F_E$ factor.

This factor evolution with respect to the incident wavelength and outer radius value can be seen in Figure 2b. Two lines following anapolar resonances (see section SIII in the Supplementary Material) are found. In particular, the more intense line at 1500 nm, corresponding to the EDA mode, displays a maximum value at $R_{Au} = 650$ nm, marking our optimum value for the first step.

After fixing the outer radius to this value, the ring inner radius is optimized in a similar manner. Figure 2c displays analogue results for the OCD average, though reversed, indicating some correlation between OCD and the amount of gold. Focusing again in the $F_E$ factor, in Figure 3b, we observe an overall maximum at $r_{Au} = 400$ nm.

After finding an optimal configuration for the structure, we characterize its THG capabilities as well as its thermal behavior, using finite element method (FEM) calculations (COMSOL Multiphysics-RF Module). Scalar third-order nonlinear susceptibilities were applied to gold ($\chi_{Au}^{(3)} = 2 \cdot 10^{-19}$ m$^2$/V$^2$) and amorphous silicon ($\chi_{a-Si}^{(3)} = 2.78 \cdot 10^{-18}$ m$^2$/V$^2$), consistent with other values found in the literature[38]. As the typical values for molecules are several orders of magnitude lower than those of gold and silicon[45], nonlinear optical effects generated within the chiral medium were neglected, thus eliminating the necessity to consider its tensorial character. Simulations were carried out in two steps, following procedures found in previous works[46,47]. In the first step, electromagnetic fields are calculated for the fundamental frequency under plane wave excitation. To do so, the background plane wave field against the substrate is calculated to serve as input for the scattering problem, which is solved for the complete geometry, considering Perfect Matched Layers (PMLs) across the boundaries of the simulation region. The second step consists in the nonlinear calculations, where the nonlinear polarization term $\boldsymbol{P}_{NL}(3\omega) = \varepsilon_0\chi^{(3)}[\boldsymbol{E}(\omega)\cdot\boldsymbol{E}(\omega)]\boldsymbol{E}(\omega)$ is taken from the fundamental frequency field calculated in the first step. This nonlinear polarization is used as a field source for the second step via an external current excitation $\boldsymbol{J}_{ext} = i3\omega\boldsymbol{P}_{NL}(3\omega)$. Additional details for the thermal analysis can be found in the Supplementary Material (section SIV) and the results are summarized in Figure 3.

Figure 3a compares the $F_E$ factor with the generated TH signal and average temperature obtained in the FEM simulations. In FEM simulations, mesh size is not uniform at any domain, and therefore spatial average values were obtained using COMSOL built-in features. A good correlation between the $F_E$ peaks and the integrated TH power $P_{TH}$ is found, except for a small $F_E$ peak at around 1100 nm, which can be associated to the ring ED resonance rather than an anapolar mode. We focus on the peaks at 1220 and 1510 nm, which are of interest as they fall within the high OCD band. The evolution of integrated TH power at those wavelengths with the pump intensity $I_0$, going from mild (0.1 mW/$\mu$m$^2$) to high values (10$^4$ mW/$\mu$m$^2$), is shown in Figure 4b. Consistently with the $F_E$ factor, the peak at 1510 nm displays higher TH powers. As seen in the figure, both peaks follow the fitted $I^3$ law, reaching efficiencies $\eta_{THG} = P_{TH}/I_0\pi R_{Au}^2$ of 0.009% (1220 nm) and 0.088% (1510 nm) at $I = 1$ GW/cm$^2$. Near field TH profiles at 1220 nm and 1510 nm are shown in the Supplementary Material (Figure S6), suggesting TH fields radiate towards the substrate at 1220 nm, and back to the source at 1510 nm.

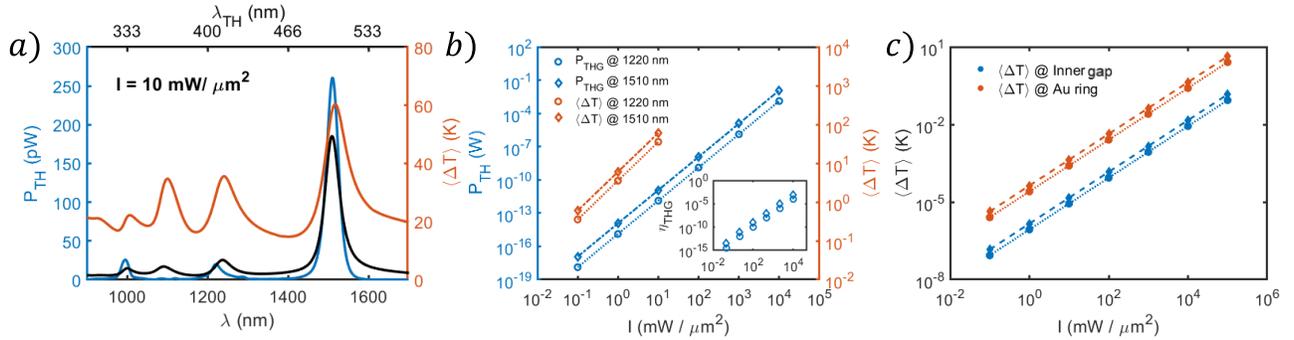

Figure 3. Results for the nonlinear and thermal analysis. a) Integrated TH and average temperature spectra compared with $F_E^3$ (black line), displaying peaks at $\lambda = 1220$ and 1510 nm. b) Evolution of TH and average temperature maxima under CW illumination for different light intensities, both fitted to cubic and linear functions respectively. The inset contains the evolution of the THG efficiency $\eta_{TH}$ with the incident intensity (circles for 1220 nm and diamonds for 1510 nm). c) Evolution of average temperature maxima (circles for 1220 nm and diamonds for 1510 nm) for different light intensities under femtosecond-pulsed illumination.

Figure 3a also shows a good correlation of the average temperature with the anapolar modes where the TH signal is being generated. High ohmic losses in metals can significantly increase temperature, potentially damaging samples. The high pump intensities required for nonlinear effects require calculating the maximum temperature rise to ensure sample integrity within a safe intensity range. For continuous wave (CW) illumination, as provided in Figure 3b, intensities up to 10 mW/$\mu$m$^2$ were considered, obtaining a maximum temperature increase of 60 K. Since temperatures above 80 ºC (corresponding approximately to the temperature increment of 60 K from room temperature) typically lead to irreversible protein denaturation[48], intensities below this value must be considered to prevent sample damage. A temperature distribution map, valid for both illumination wavelengths, is shown in the Supplementary Material (Figure S6).

Although THG usually requires high illumination intensities, and CW illumination has a relatively low intensity threshold for thermal damage, THG typically uses femtosecond lasers with microsecond repetition times, exceeding the cooling time of typical gold nanoparticles in water[49]. This allows heat dissipation into the high thermal conductivity surrounding media, preventing overheating. Following previous literature, time-dependent heat transfer simulations were carried out in COMSOL assuming typical femtosecond illuminations. The results can be seen in Figure 3c, showing the evolution of maximum temperatures with the fluence/intensity. Here, the average temperature increments still follow a linear dependence with fluence/intensity, but the peak values are much lower, and for all considered intensities (up to 10 GW/cm$^2$) no thermal damage is expected.

Finally, to demonstrate the practical functioning of our design, we propose a metasurface design to further illustrate its capabilities to enhance the CD effect. After an optimization process (Supplementary Material section SV), we set up a square array with period $\Lambda = 1850$ nm and cover it with a chiral thin layer, representing a sample, with thickness $\delta = 20$ nm, characterized by a constant refractive index $n = 1.45 - 0.01i$ and a realistic complex scalar Pasteur parameter $\kappa = 10^{-4} + 10^{-6}i$, owing to the random orientation of molecules[1,50]. Details for the chiral FEM simulations can be found in the Supplementary Material (section SVI). Linear and nonlinear CD signals were calculated as

$$CD = \operatorname{atan}\left(\frac{T_R - T_L}{T_R + T_L}\right) ; \quad CD_{TH} = \operatorname{atan}\left(\frac{P_R^{3\omega} - P_L^{3\omega}}{P_R^{3\omega} + P_L^{3\omega}}\right) \qquad (4)$$

where $T_{R,L}$ represents transmittance for either right or left-handed incident polarizations, and $P_{R,L}^{3\omega}$ is the integrated TH power (either in reflection or transmission directions) for the same incidence. Since the CD effect is correlated to the volume of chiral material[34], we increase the radius of the inner gap to $r_{Si} = 50$ nm to place more molecular volume in the high chirality inner gap.

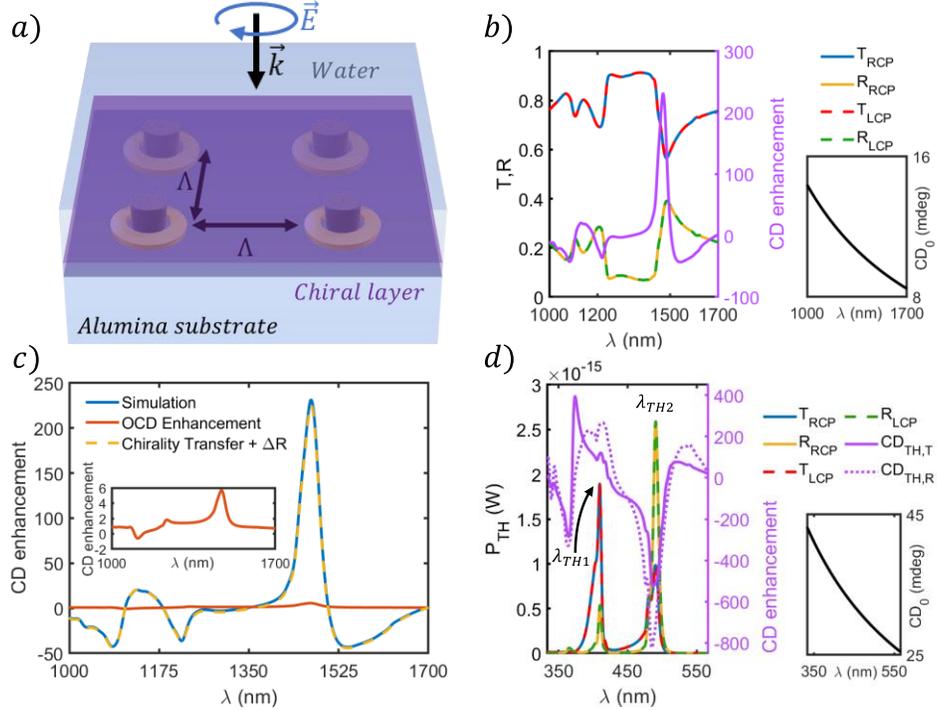

Figure 4. a) Scheme of the proposed metasurface. b) Transmittance and reflectance spectra for the simulated metasurface with RCP and LCP illuminations, as well as the obtained transmission CD enhancement with respect to a bare chiral layer, whose CD signal is shown in the inset. c) Simulated CD enhancement compared with the theoretical value predicted by OCD enhancement and the contribution from the chirality transfer mechanism. d) TH integrated power spectra. $T$ refers to integration below the metasurface and $R$ refers to power integrated above the metasurface. The CD enhancement is shown for both integrations, compared to the transmission CD spectrum of a bare chiral layer at TH frequencies.

The results for these calculations are summarized in Figure 4, with the proposed metasurface depicted in Figure 4a. Figure 4b contains the calculated linear reflectance and transmittance for both polarizations, as well as the CD effect enhancement $CD_{Enh} = CD/CD_0$ with CD calculated from equation 4 and $CD_0$ being the CD signal from a bare chiral layer covering the substrate, which is shown in the corner inset. The $CD_0$ signal was calculated analytically using a Transfer Matrix Method approach[50]. Two peaks (one negative and one positive) can be observed, corresponding to the anapole modes, at both ends of the high transmittance band, with the higher one reaching approximately a 230-fold enhancement.

Figure 4c compares the obtained simulation results with the theoretical values given by $CD_{Enh,OCD} = \frac{\langle C \rangle/C_{CPL}}{T}$ [29] and the contribution from the chirality transfer mechanism (differential absorption in the nanostructure as a result of scattering from the chiral medium due to its real part of $\kappa$) plus other effects from differential reflection[51]. Here, due to the relationship between the real and imaginary parts of $\kappa$ ($10^2$ across all the spectrum), chirality transfer dominates the shape of the signal, particularly at the observed peaks, which correspond to anapole resonances. This can be explained as the field confinement caused by anapole resonances enhances the light-matter interaction in the chiral medium, significantly strengthening the effect. Further discussion, including additional simulation results with a pure imaginary Pasteur parameter and its agreement with theory can be found in the Supplementary Material (section SVII).

Finally, Figure 4d shows the results for the conducted nonlinear simulations, including integrated powers above and below the metasurface (similar to reflection and transmittance) for both circular polarizations, as well as the CD effect enhancement. At the first THG wavelength ($\lambda_{TH1} = 410$ nm), transmission dominates for both polarizations, whereas at the second THG wavelength ($\lambda_{TH2} = 490$ nm), reflection prevails. In this case, we find that the CD effect is significantly enhanced at both wavelengths, with 100-fold enhancements compared to the bare chiral film and beyond. In particular, the second THG wavelength showcases an 800-fold enhancement in reflection, suggesting the utility of this metasurface in both transmission (linear CD) and reflection (nonlinear CD) applications.

In conclusion, we have presented a hybrid structure consisting on a holed silicon nanodisk surrounded by a gold ring, achieving high average OCD and efficient THG. Using FDTD simulations and multipole analysis, we related the high OCD with the spectral overlap of electric and magnetic resonances within the disk's gap, including anapoles. Thermal calculations confirmed no thermal side effects under pulsed illuminations.

To give more insights on the practical functioning of our design, FEM simulations with chiral matter mimicking a real sample were carried out. The proposed metasurface showed up to 230-fold enhancement of the CD effect in transmittance for the linear regime and an 800-fold enhancement in reflectance for the nonlinear TH regime, with a good agreement with theory. We believe that these results lay the foundations towards enhanced measurements of molecular chirality.

## SUPPLEMENTARY MATERIAL

The supplementary material contains extra notes and figures to strengthen the claims made in this paper. In particular, results for the scattering cross-sections of different aspect ratio silicon holed disks in water with corresponding multipole decompositions, an analysis on the influence of the alumina substrate on the average optical chirality density and multipole analysis, details for the geometrical optimization of the gold ring and multipole analysis, details on thermal FEM calculations, results for the Third Harmonic and temperature profiles, the optimization of the metasurface, details on chiral FEM calculations and a complementary analysis on the chirality transfer mechanism and the differential transmittance and reflectance in metasurfaces, showing compatibility between theory and simulation results.

## ACKNOWLEDGEMENTS

This work acknowledges funding from the MOPHOSYS Project (PID2022-139560NB-I00) from Proyectos de Generación de Conocimiento provided by the Spanish Agencia Estatal de Investigación. G. S. thanks the Spanish Ministry of Education for his predoctoral contract grant (FPU21/02296). J.G.-C. thanks the Ministry of Science and Innovation of Spain for his FPI grant (PRE2019-088809).

## AUTHOR DECLARATIONS

### Conflict of interest

The authors have no conflict to disclose.

### Author contributions

**Guillermo Serrera**: Conceptualization (supporting); Investigation (lead); Formal analysis; Methodology (equal); Writing – original draft (lead); Writing – review & editing (equal). **Javier González-Colsa**: Investigation (supporting); Methodology (equal), Writing – original draft (supporting); Writing – review & editing (equal). **Pablo Albella**: Conceptualization (lead); Funding acquisition; Resources; Supervision; Writing – review & editing (equal).

# Amplified linear and nonlinear chiral sensing assisted by anapole modes in hybrid metasurfaces

Guillermo Serrera, Javier Gonzalez-Colsa and Pablo Albella[a]
**AFFILIATION**
*Group of Optics, Department of Applied Physics, University of Cantabria, 39005, Spain.*
a) Author to whom correspondence should be addressed: pablo.albella@unican.es

# Supplementary Material

## SI. SCATTERING CROSS-SECTIONS OF A SILICON HOLED DISK IN WATER

The results of the multipole decomposition analysis carried out for the results shown in Figure 1 of the main document are shown in Figure S1 and corresponding to the 0.1 steps of the interval $\frac{h}{R_{Si}} \in [0.5, 1.7]$. An almost perfect fit to the FDTD results is achieved upon consideration of multipoles up to octupole order. The categorization of resonances in Figure 1a is based on peak recognition for ED, MD, EQ and MQ resonances, while anapolar modes (EDAs, MDAs and EQA) were recognized as valleys.

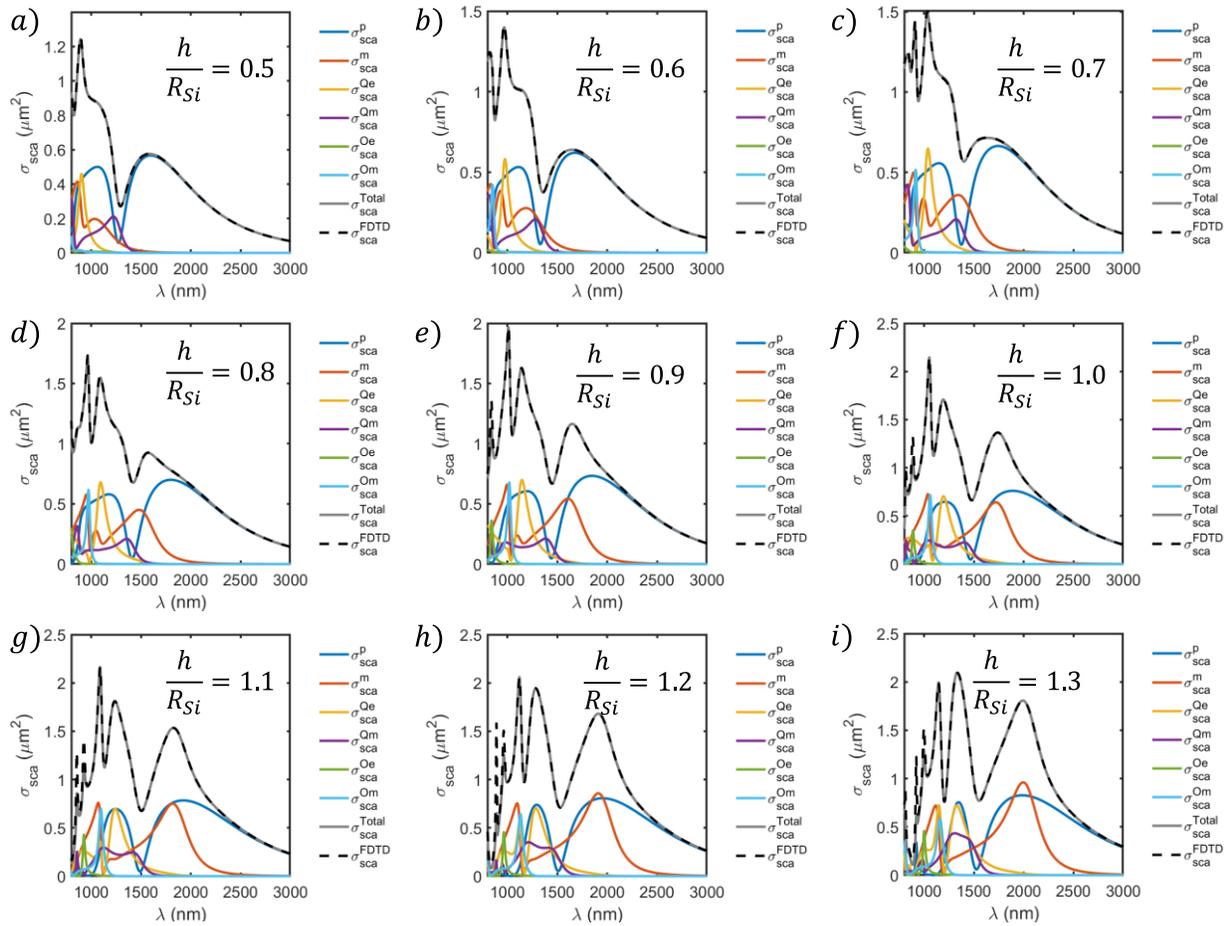

Figure S1. Scattering cross-section with multipole decomposition for silicon holed disks with aspect ratios a) 0.5 to m) 1.7.

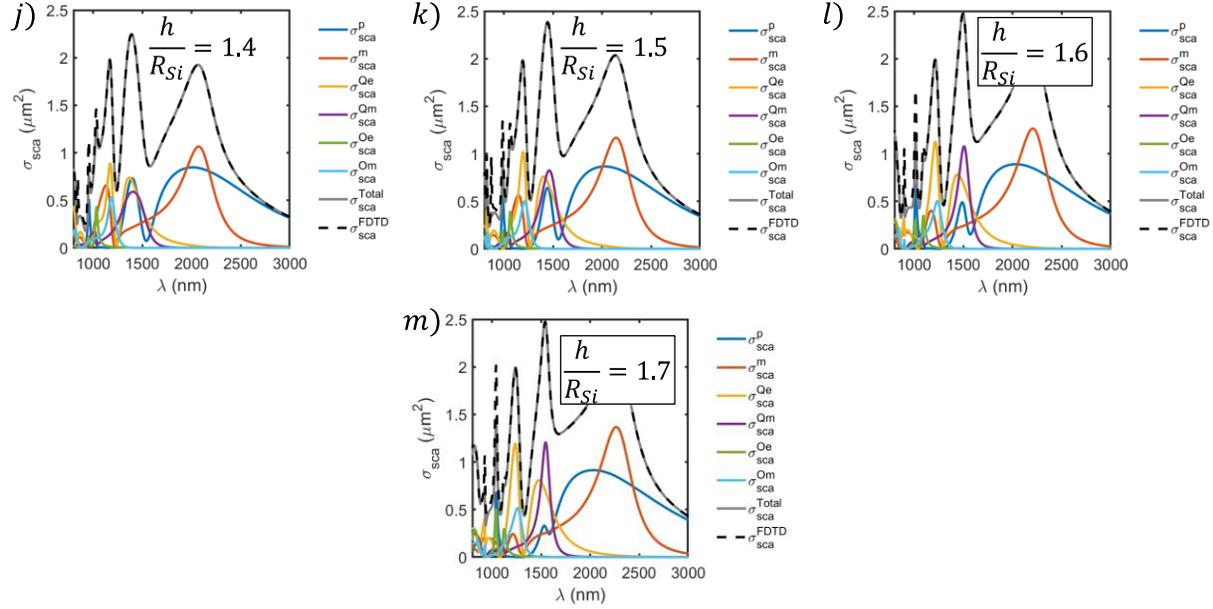

Figure S1 (cont.). Scattering cross-section with multipole decomposition for silicon holed disks with aspect ratios a) 0.5 to m) 1.7.

## SII. INFLUENCE OF THE ALUMINA SUBSTRATE ON THE AVERAGE OPTICAL CHIRALITY DENSITY

The results for averaged optical chirality density, electric field, magnetic field and phase factor are summarized in Figure S2 in a similar format than Figure 1 of the main text. We observe that, in general, fields are less intense, leading to a lower average chirality within the disk gap. However, all trends remain the same and the phase factor is not affected in any way.

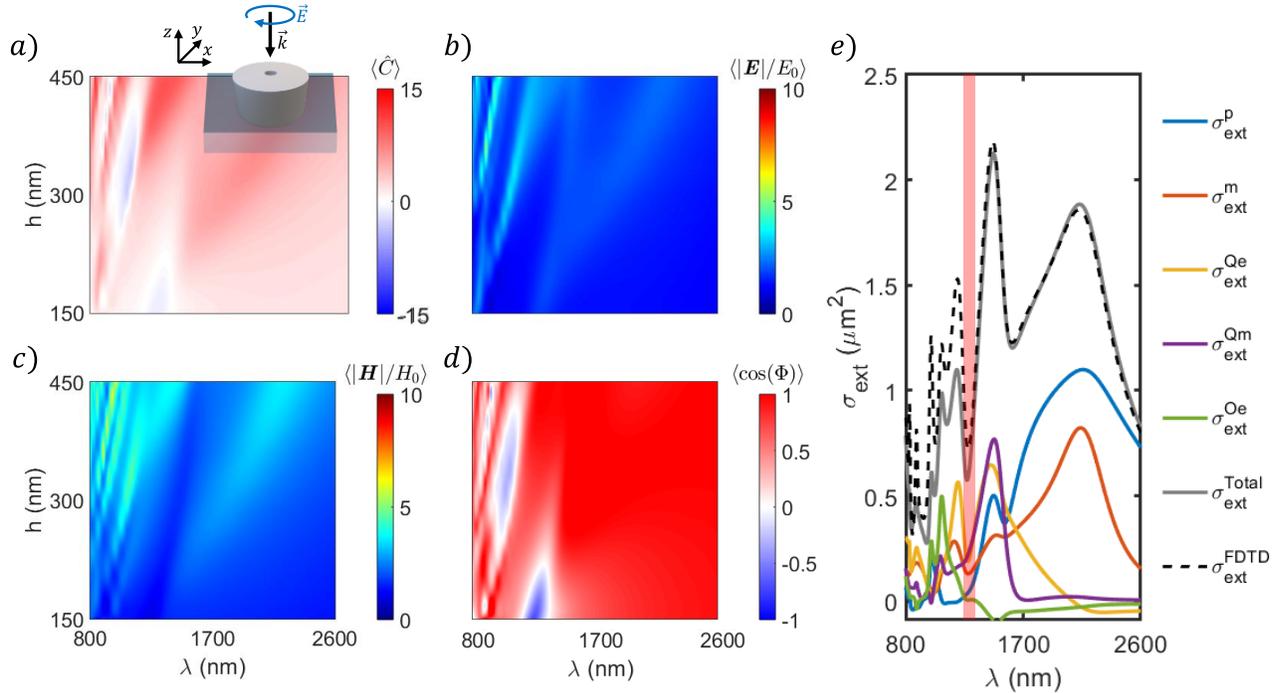

Figure S2. Evolution of the a) optical chirality density, b) electric field, and c) magnetic field enhancements and d) phase factor $\cos(\Phi)$ averaged within the inner gap of the disk-substrate system with respect to the disk height and the incident wavelength $\lambda$. e) Extinction cross-section for the $h = 450$ nm case with multipole decomposition up to the electric octupole term. In red, the spectral position of point A, where anapole resonances overlap, is highlighted.

This is also seen in the far field multipole decomposition of the extinction cross-section (Fig. S2e), obtained following work by Evlyukhin and coworkers[1]. In this case, precision in comparison to FDTD results is lower as a substrate introduces cross-talk between multipoles[2], meaning that the considered multipoles are not the true eigenmodes of the system[1]. This also explains the fact that some multipoles have a negative contribution into the cross-section. Despite this, the same trends are generally observed, as the substrate has a relatively low refractive index contrast with water. As often happens with substrates, the electric dipole contribution is widened due to the mode profile extending into the substrate[1]. A notorious dark mode near $\lambda = 1300$ nm is found as a result of several anapole resonances (mainly magnetic dipole and electric quadrupole) overlapping.

## SIII. GEOMETRICAL OPTIMIZATION OF THE GOLD RING

As mentioned in the main text, the optimization process consisted of two steps. In the first one, the inner radius was fixed to the minimum considered value ($r_{Au} = 350$ nm), while the outer radius was varied until an optimum point was found. The results for this first step, together with more details, are found in Figure S3.

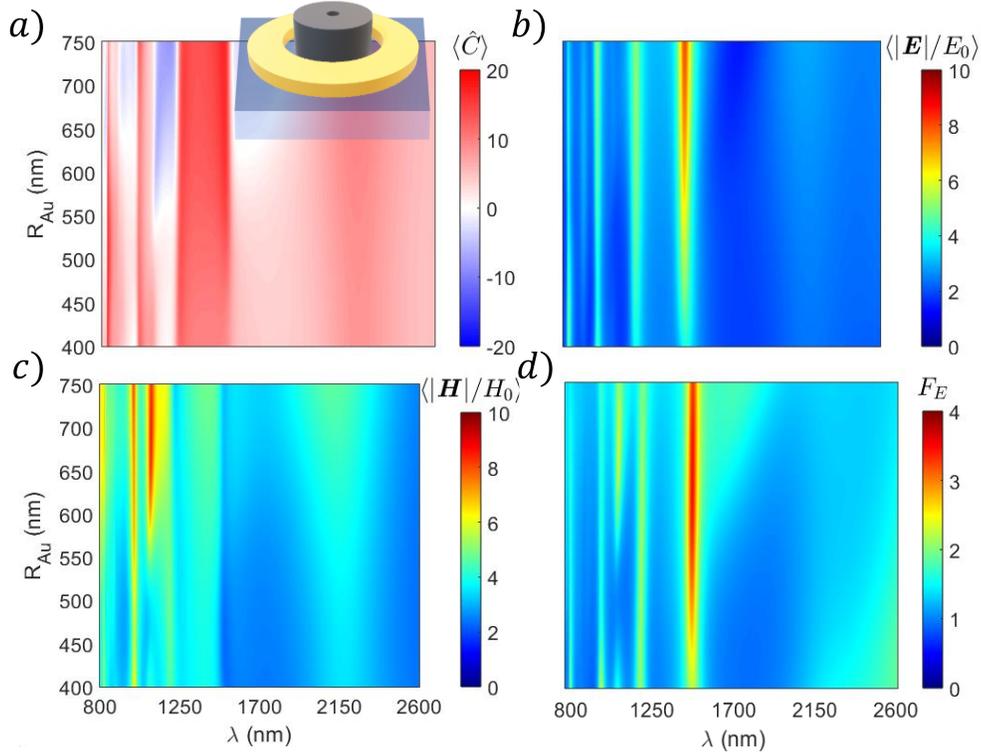

Figure S3. FDTD results obtained for the optimization of the gold ring. a) Results for the optical chirality density, b) normalized electric field intensity $F_E$ inside the silicon disk, c) electric field and d) magnetic field enhancements; for several outer radii of the gold ring, with a fixed inner radius $r_{Au} = 350$ nm.

Figures S3a and S3d are equivalent to Figure 2a and 2b in the main text and are repeated here for reference. Focusing on the extra information provided by this figure, we can see that near the chirality enhancing broadband, two narrow lines with intense local electric fields appear (Figure S3b). These correspond to the EDA mode (around 1500 nm) and the superposition of modes discussed in the previous section (around 1220 nm). This is further confirmed by the magnetic field map (Figure S3c), where we find a more broadband amplification, contributed by the MQ mode (near the EDA resonance); and a narrow line corresponding to the MDA mode near 1200 nm. In general, all fields seem to have larger intensifications for bigger gold rings (larger radii).

Since the OCD average in the 1230-1500 nm band does not vary significantly with respect to $R_{Au}$, and the $F_E$ factor reaches it maximum at $R_{Au} = 650$ nm, this value was fixed for the second step of the optimization process, where the inner radius was varied. The results for this step are found in Figure S4. Again, Figures S4a and S4d are equivalent to Figure 2c and 2d of the main tex. In general, this figure displays analogue results to the previous one. The same trends found in the first step are observed, though reversed, suggesting that the overall field concentrations are correlated with the amount of gold in the structure. Focusing again in the $F_E$ factor, in Figure S4d, we observe an overall maximum at $r_{Au} = 400$ nm.

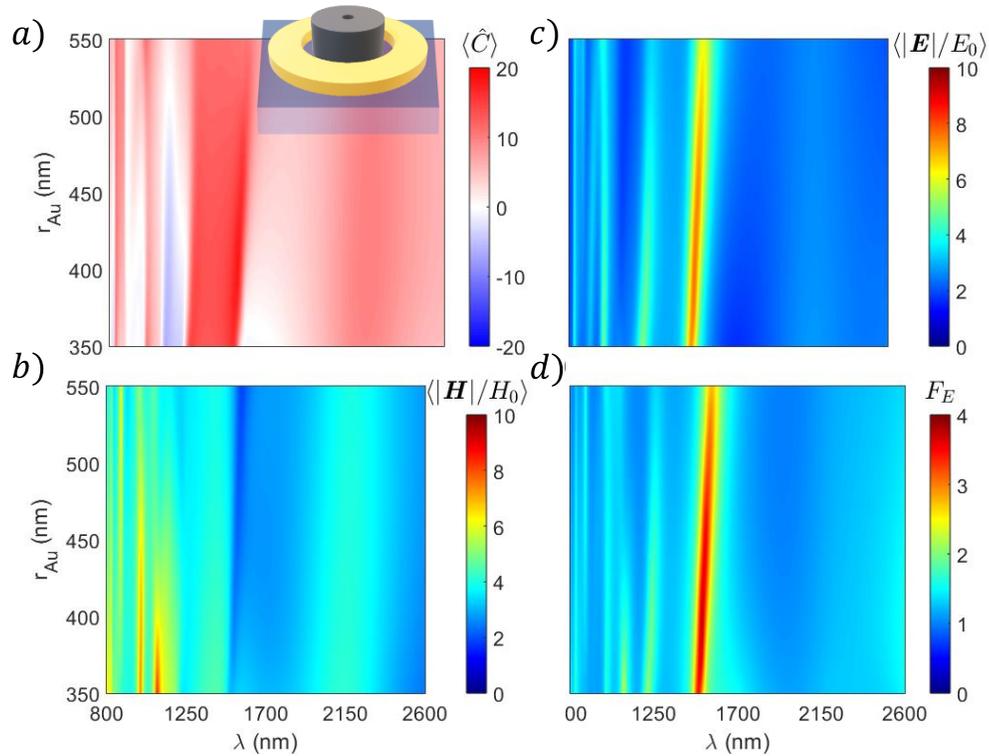

Figure S4. FDTD results obtained for the optimization of the gold ring. a) Results for the optical chirality density, b) normalized electric field intensity $F_E$ inside the silicon disk, c) electric field and d) magnetic field enhancements; for several inner radii of the gold ring, with a fixed outer radius $R_{Au} = 650$ nm.

To obtain a better understanding of the electromagnetic response, we perform a multipole decomposition of the gold-ring system optimal geometry in an homogeneous water medium. As demonstrated in Figure S2e, the addition of the substrate does not significantly change the overall response of the system in comparison to the homogeneous result in Figure 1e of the main text. However, the addition of the substrate causes a significant limitation to the precision of the multipole decomposition method. Therefore, the substrate has been eliminated from this calculation for the sake of clarity.

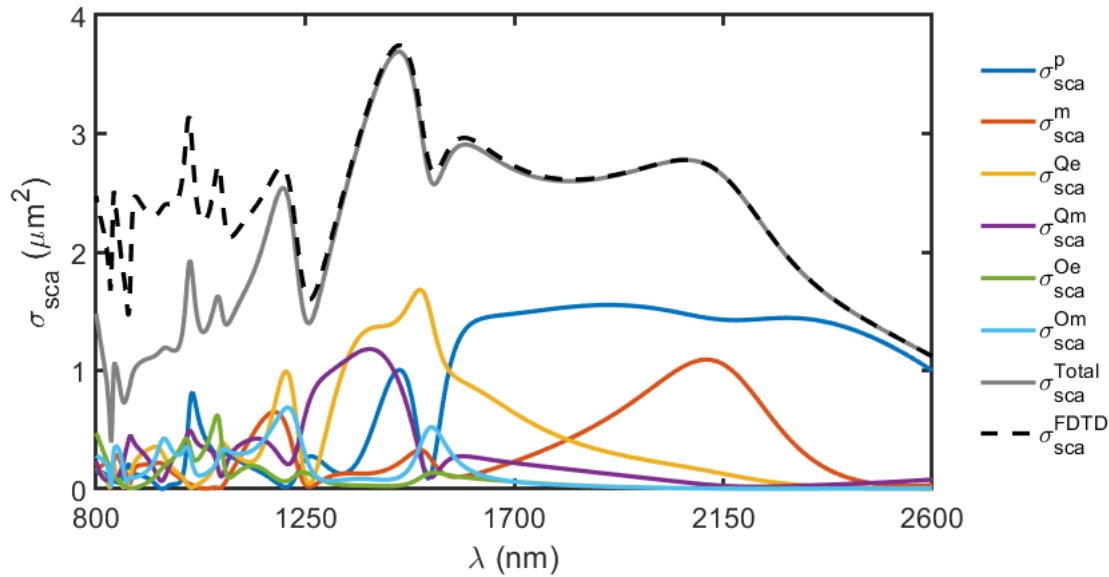

Figure S5. Far field scattering cross-section for the $(R_{Au}, r_{Au}) = (650, 400)$ nm system with multipole decomposition up to the magnetic octupole term.

The results for this analysis are shown in Figure S5. Here we can see that a good fit with FDTD is obtained for wavelengths longer than 1200 nm. A very broad electric dipolar response is seen at the longer wavelengths, owing to the well-known broad response of gold rings[3,4]. The introduction of the ring also redshifts the excitation of higher order terms significantly, as the electric quadrupole displays a dual maximum at 1400 and 1500 nm. The magnetic octupole resonance is also redshifted to 1500 nm. This general redshift also explains the mismatch with FDTD at shorter wavelengths since contributions from higher terms become important. Despite all this, the dark modes are still observed, near 1250 (main contributions from magnetic dipole and electric quadrupole terms) and 1500 nm (electric dipole contribution) respectively. These agree well with the results in Figures S3 and S4, especially considering the spectral position of $F_E$ maxima; and support the idea that an efficient THG is obtained due to anapole resonances.

## SIV. DETAILS OF THERMAL AND NONLINEAR CALCULATIONS

In order to obtain precise results for thermal calculations, we use the RF module of the COMSOL Multiphysics suite since it provides with a built-in strategy to solve both light-matter interaction and thermal dissipation problems in an integrated manner. To keep consistency with the previously conducted FDTD simulations, material data was extracted and interpolated from the same sources.

Analogously to non-linear calculations, the light-matter interaction problem under CW illumination was solved in two steps by means of FEM. Initially, Maxwell's equations are solved to determine the background electric field (plane wave) within the simulation region in absence of the plasmonic structure. Subsequently, these equations are solved again, this time considering the nanostructure boundary conditions. This step provides with the spatial distribution of the scattered electric field, the scattering and absorption cross sections, and the absorbed/emitted electromagnetic power, among others. The second step involves solving the heat equation, based on the previous electromagnetic analysis. Particularly, the absorbed electromagnetic power (resistive losses) is used to feed the thermal calculations which finally allows for obtaining the spatial temperature fields. A convective node across the simulation region boundaries was used to enable the cooling of the whole system.

The near field spatial distributions of TH electric field and temperature are shown in Figure S6. Note that, once normalized, the temperature distribution is the same for all wavelengths. We observe in Figures S6a-b, that the TH field suggests transmission for 1220 nm incidence, while emission in the 503 nm (1510 nm incidence) wavelength is mostly side and upwards.

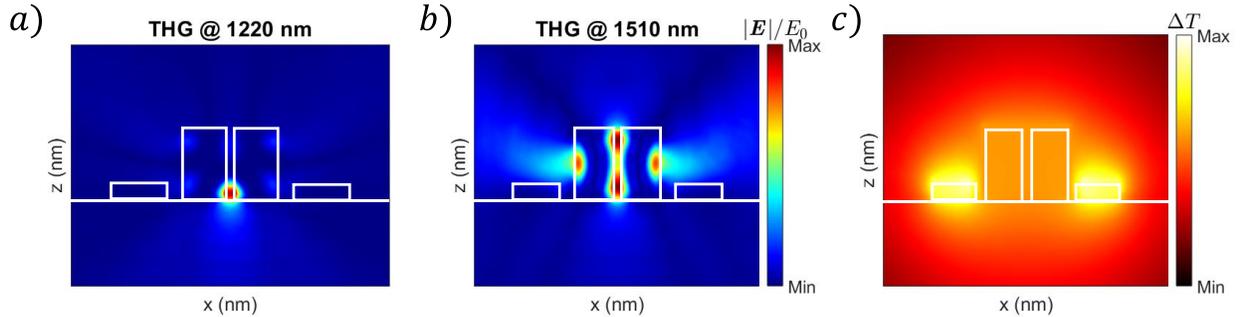

Figure S6. Normalized spatial distributions of a) TH electric field at a 406 nm wavelength (1220 nm incidence), b) TH field at 503 nm (1510 nm incidence) and c) temperature increment.

To explore the thermal behavior of the system under pulsed illumination, we adopt the theoretical framework established by Baffou and coworkers[5], assuming a Gaussian temporal profile with peak intensity $I_0$ equivalent to the intensities studied in Figure 3b; and a pulse length of 100 fs. This pulse length is sufficiently shorter than the characteristic gold electron phonon interaction time ($\tau_{e-ph} = 1.7$ ps), so that the heat power can be properly approximated as $Q(t) \approx \sigma_{abs} F p(t)$, where $\sigma_{abs}$ is the wavelength-dependent absorption cross-section of the gold ring, obtained from previous CW simulations, $F$ is the pulse fluence and $p(t) = \frac{1}{\tau_{e-ph}} \exp\left(-\frac{t}{\tau_{e-ph}}\right)$ the time-dependent power transfer function from the electron gas to the atomic lattice. Interfacial Thermal Conductances (ITCs) were considered at the gold-water (200 MW/m$^2\cdot$K [6]) and silicon-water (143 MW/m$^2\cdot$K [7]) interfaces.

FEM time-dependent heat transfer simulations were carried out in COMSOL, covering a whole microsecond, which is the time between pulses for an assumed repetition rate of 1 MHz. To show that this repetition rate is sufficient to allow thermal dissipation between pulses, Figure S7 contains the temporal evolution of temperature in both the gold ring and the inner gap of the silicon disk.

It can be seen that the thermal response of the structure is not immediate and temperature increments take around 100 fs to rise, reaching maximum values around the picosecond range and then fall, vanishing at $t_{rep} = 1\ \mu s$, meaning that for the assumed pulse repetition rate $1/t_{rep} = 10^6$ Hz the structure will not accumulate heat between pulses. For the inner gap of the structure, where samples are intended to be located, this growth and decay shape of the temperature increment is significantly delayed, due to the heat transfer from the gold ring and the overall lower thermal conductivity of water compared to that of the alumina substrate. Likewise, it reaches zero values at the end of the pulse cycle suggesting negligible potential overheating of the sample.

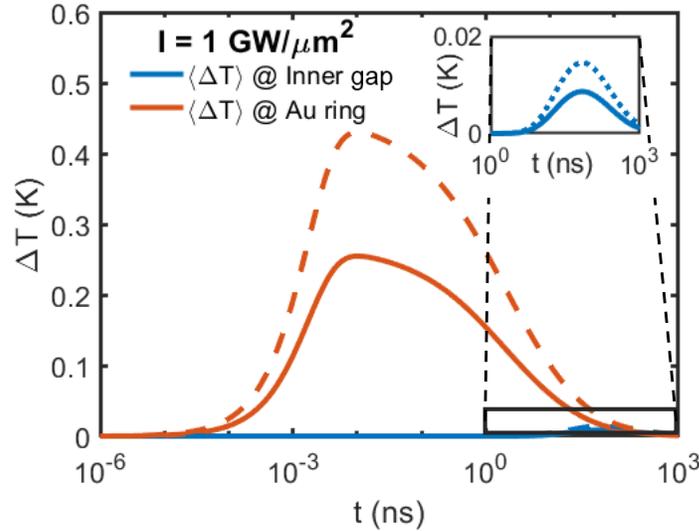

Figure S7. Time evolution of the average temperature inside the inner gap of the Si disk (blue) and inside the gold ring (red) from $t = 1$ ns to $t = 1\ \mu s$ at 1220 nm (solid lines) and 1510 nm (dashed lines) wavelengths.

## SV. METASURFACE OPTIMIZATION

Once the unit cell has been designed and analyzed, we propose a metasurface design to further illustrate its capabilities to enhance the CD effect. A square lattice with period $\Lambda$ is arranged, and the dependence of OCD with this period is studied for the interval $\Lambda \in [1500,3000]$ nm in steps of 100 nm. Both the transmittance and reflectance of the metasurface, as well as the average OCD inside the inner gap were calculated using FDTD methods, with the relevant results shown in Figure S8.

As shown in Figure S8a, the metasurface period does not have a significant impact on the spectral position of the high OCD band, being approximately constant in the 1230-1500 nm interval for both tightly packed ($\Lambda = 1500$ nm) and widely spaced ($\Lambda = 3000$ nm) metasurfaces, where the overall behaviour is close to that of the unit cell. This can be explained from the period length, which is higher than the high OCD band wavelengths, meaning that coupling between unit cells will be generally weak. However, it can also be seen that despite the homogeneity in the band's spectral position, there are significant inhomogeneities in the actual average OCD value along that band. For this reason, in Figure S7b the spectrally averaged OCD values within that band are plotted against the lattice period, with a visible maximum at $\Lambda = 1850$ nm. For this lattice period, the reflectance and transmittance spectra are represented in Figure S7c. In general, these spectra show high transmittances for all wavelengths, reaching values around 0.9 for the [1250,1400] nm interval, and minima around 0.6-0.7 for the wavelengths where the anapolar modes are expected. This characteristic is also interesting since high transmittances, while not beneficial for CD enhancement, can lead to more precise measurements in practice[8].

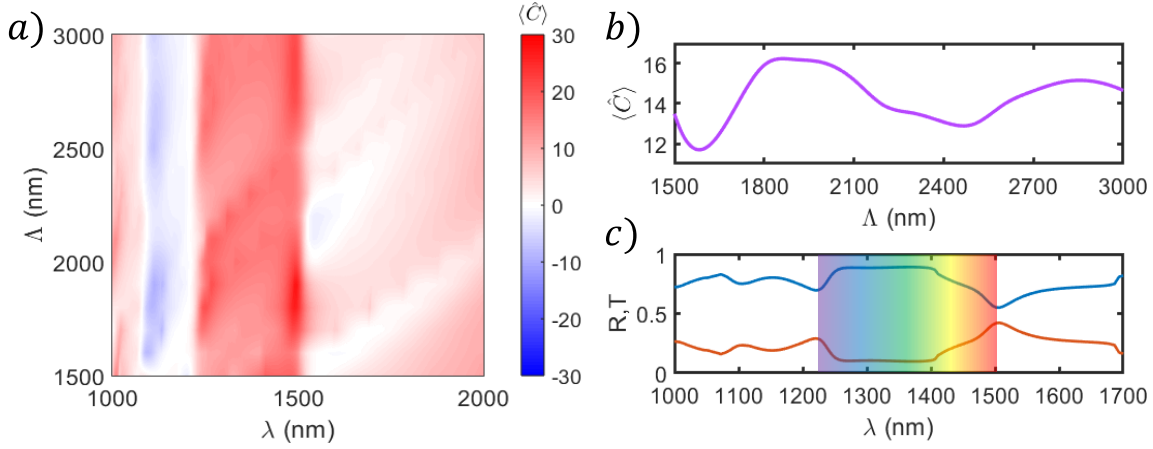

Figure S8. Average OCD in the inner gap of the structure. a) Calculated vs incident wavelength $\lambda$ and lattice period $\Lambda$. b) Averaged both spatially (within the inner gap) and spectrally (in the 1200-1500 nm band) vs the lattice period $\Lambda$. c) Reflectance (orange) and transmittance (blue) spectrum for the optimal $\Lambda = 1850$ nm metasurface design, with the high OCD enhancement band in the background.

## SVI. DETAILS OF CHIRAL FEM SIMULATIONS

As mentioned in the main text, the electromagnetic interaction of chiral media is mediated by the Pasteur parameter in the constitutive relations

$$\begin{aligned}\boldsymbol{D} &= \varepsilon\boldsymbol{E} - i\kappa\sqrt{\varepsilon_0\mu_0}\boldsymbol{H} \\ \boldsymbol{B} &= \mu\boldsymbol{H} + i\kappa\sqrt{\varepsilon_0\mu_0}\boldsymbol{E}\end{aligned} \quad (S1)$$

To perform numerical calculations, the built-in constitutive relations in COMSOL were effectively transformed into equations S1. The chiral medium, mimicking a real sample, was set to a constant refractive index $n = 1.45 - 0.01i$ and a complex Pasteur parameter $\kappa = 10^{-4} + 10^{-6}i$, exhibiting a large but still realistic magnitude, with a typical ratio between real and imaginary part[9]. To minimize numerical error, mirror-symmetric meshing in the simulation region was utilized[10].

## SVII. CIRCULAR DICHROISM AND CHIRALITY TRANSFER: THEORY AND SIMULATION

Simulation results in Figure 4 in the main text showcase that CD transmission enhancement is not proportional to the average OCD enhancement divided by the mean transmittance as predicted in other works[11]. This equation holds when the differential reflectance in the metasurface $\Delta R$ is neglected in comparison with the differential transmittance $\Delta T$. Then, a relation between $\Delta T$ and the differential absorbance $\Delta A$ can be established as

$$\Delta T = -\Delta A - \Delta R \simeq -\Delta A \quad (S2)$$

where the differential absorbance can then be related to the chirality transfer mechanism and the OCD enhancement divided by the incident power[12]

$$CD_{Enh} = \frac{CD}{CD_0} = CD_{Enh,CT} + CD_{Enh,OCD} = \frac{\Delta P_{abs}/\Delta P_{abs,0}}{T} + \frac{\langle C \rangle / C_{inc}}{T} \quad (S3)$$

However, as Figure S9a showcases, differential reflectance and transmittance are similar in absolute value for all wavelengths, meaning that the contributions from OCD enhancement and chirality transfer can not be established following equation S3. In order to establish this contribution, additional FEM simulations were carried out, with a pure imaginary Pasteur parameter $\kappa = 10^{-3}i$, consistent with the optical modelling of chiral media in other works[11,13,14].

These results can be found in Figure S9b and c, which show that CD enhancement by simulation is far more similar to the OCD enhancement term in equation S3. Slight mismatches in amplitude and wavelength are found between theory and simulation. These can be attributed to substantial differential reflectance near the CD enhancement maximum.

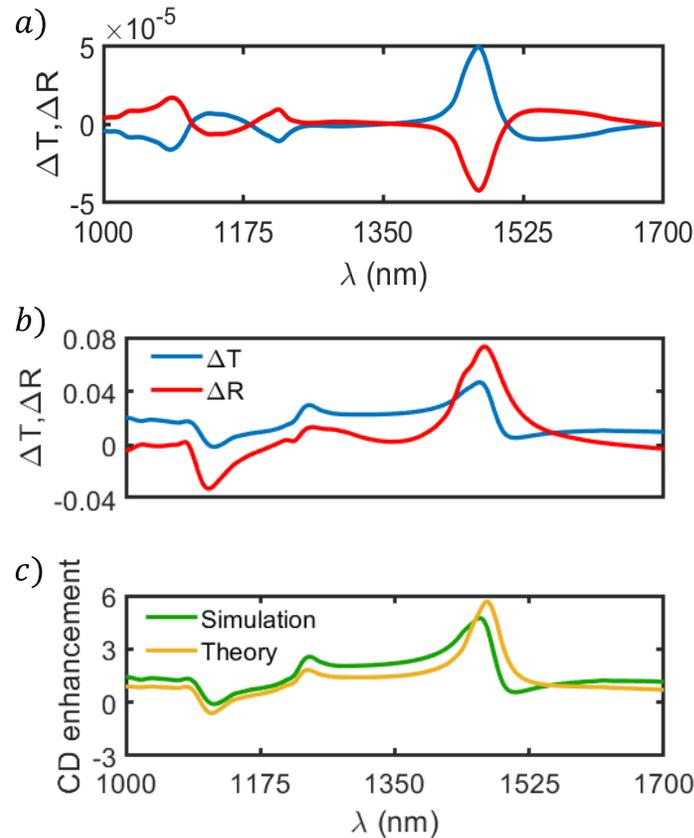

Figure S9. Differential transmittance ($\Delta T$) and reflectance ($\Delta R$) between circular polarizations in the $\Lambda = 1850$ nm metasurface with chiral medium characterized by a) $\kappa = 10^{-4} + 10^{-6} i$, and b) $\kappa = 10^{-3} i$. c) Simulation and theory results for CD enhancement with $\kappa = 10^{-3}$.

Since for the system with $\kappa = 10^{-4} + 10^{-6} i$ the terms in equation S3 cannot be explicitly separated due to the significant differential reflectance, chirality transfer contribution in Figure 4c was put together with $\Delta R$ as the difference between the simulation result and the OCD enhancement term.